\documentclass[twocolumn,showpacs]{revtex4}

\usepackage{graphicx}
\usepackage{dcolumn}
\usepackage{amsmath}
\usepackage{epsfig}

\makeatletter
\def\btt#1{\texttt{\@backslashchar#1}}%
\DeclareRobustCommand\bblash{\btt{\@backslashchar}}%
\makeatother

\begin{document}

\preprint{HEP/123-qed}

\title[Short Title]{The spin-$\frac{1}{2}$ $J_1$-$J_2$ model on the
body-centered cubic 
  lattice} 

\author{R. Schmidt}
 \email{reimar.schmidt@physik.uni-magdeburg.de}
\author{J. Schulenburg}
\author{J. Richter}
\affiliation{
Institut f\"ur Theoretische Physik,
         Universit\"at Magdeburg,
         P.O.Box 4120, D-39016 Magdeburg, Germany
}

\author{D. D. Betts}
\affiliation{
Department of Physics, Dalhousie University, 
         Halifax, N.S., Canada, B3H 3J5
}%

\date{\today}

\begin{abstract}
Using exact diagonalization (ED) and linear spin wave theory (LSWT)
we study the influence of frustration and quantum fluctuations on the
magnetic  ordering in the ground state of the 
spin-$\frac{1}{2}$ $J_1$-$J_2$ Heisenberg antiferromagnet ($J_1$-$J_2$ model)
on the body-centered cubic (bcc) lattice.  
Contrary to the $J_1$-$J_2$ model on the square lattice, we find for the bcc
lattice that frustration and 
quantum fluctuations do not lead to a quantum 
disordered phase for strong frustration. 
The results of both approaches (ED, LSWT)
suggest  a first order
transition at $J_2 /J_1 \approx 0.7$
from the  two-sublattice N\'eel
phase at low $J_2$  to a collinear phase at large $J_2$.
\end{abstract}

\pacs{Valid PACS appear here}
\maketitle


\section{Introduction}
\label{intro}

The properties of the two-dimensional (2d) $J_1$-$J_2$ model have attracted a 
great deal of interest during the last decade 
(see e.g. Refs. \cite{squa_ref7,squa_ref8,squa_ref9,zfp93,
squa_ref9a,squa_ref10,squa_ref11,squa_ref12,Sushkov01,squa_ref13,squa_ref13a} 
and references therein).
The Hamiltonian of the $J_1$-$J_2$ model is
\begin{eqnarray}
\label{ham}
H=J_1 \sum_{\langle i,j \rangle}{\bf S}_i \cdot {\bf S}_j
+J_2 \sum_{[ i,j ]}{\bf S}_i \cdot {\bf S}_j ,
\end{eqnarray}
where $J_1=1$ is the nearest-neighbor, 
and $J_2 \ge 0$ is the frustrating next-nearest-neighbor Heisenberg exchange. 
We are interested in the extreme quantum case, i.e. we consider 
the spin quantum number is $s =1/2$. 
For the square lattice it seems to be well-accepted that
there is a quantum spin-liquid phase between 
$J_2 /J_1 \approx 0.38$ and 
$J_2 /J_1 \approx 0.60$ and that the corresponding 
quantum phase transitions from the 
N\'eel ordered state to the spin-liquid state at 
$J_2 /J_1 \approx 0.38$ is
of second order. The nature of the  transition from the spin-liquid state to 
the  collinear state at $J_2 /J_1 \approx 0.60$ is still under discussion
but there are indications that it might be of  first order 
\cite{Sushkov01}. 
The N\'eel phase for small $J_2$ is characterized by 
an antiparallel alignment of
nearest-neighbor spins with a corresponding  magnetic wave vector  
  ${\bf Q}_{N\acute{e}el}=(\pi ,\pi )$.
The collinear state for large  $J_2$ is twofold degenerated
and the corresponding magnetic wave vectors are   ${\bf Q}^1_{col} =(\pi , 0)$
and ${\bf Q}^2_{col} =(0 ,\pi )$. 
The two collinear states are characterized by a parallel spin orientation of 
nearest neighbors in vertical (horizontal) direction and an antiparallel 
spin orientation of  nearest neighbors in horizontal (vertical) 
direction and exhibit therefore 
N\'eel order within the
initial sublattices $A$ and $B$. 
The properties of the  spin-liquid phase are  
a current field of active research. Even additional quantum phase transitions 
at $(J_2 /J_1 )\approx 0.34$
and $(J_2 /J_1 )\approx 0.50$ are discussed
\cite{Sushkov01}.

The properties of quantum spin systems strongly depend on the
dimensionality.
So, contrary to the 2d model, 
the one-dimensional $J_1-J_2$ model
does not have a N\'eel ordered ground state, but
exhibits a transition from a critical state to a dimer phase at
$J_2/J_1=0.241$ (see e.g. \cite{1d_a,1d_b,1d_c}). 
Though the tendency to order is
more pronounced in three-dimensional (3d)
quantum  spin systems than in low-dimensional ones a spin-liquid phase is
also
observed for frustrated 3d systems like the Heisenberg antiferromagnet on
the pyrochlore lattice \cite{canals98,koga01}. 

In this paper we consider the 3d version 
of the $J_1-J_2$ model. To cover
the possibility to have in a 3d model both N\'eel phases  
found for the 2d model we need a 
3d bi-bipartite
lattice, i.e. a lattice consisting of two interpenetrating bipartite
sublattices.   
The bcc lattice consists of two interpenetrating, identical simple cubic
sublattices. Each simple cubic sublattice consists of two interpenetrating,
identical fcc lattices. Therefore the bcc lattice is a 3d bi-bipartite cubic
lattice. 

The classical ground state for the bcc lattice
corresponds to that of the square lattice: For $J_2/J_1 < \alpha_c$ it is a
usual two-sublattice N\'eel state whereas for $J_2/J_1 > \alpha_c$   
an antiferromagnet with four sublattices is realized. 
The transition point $\alpha_c$ 
depends on the coordination numbers $\alpha_c={z_1}/({2z_2})$, 
where $z_1$ is the number of nearest-neighbors ($J_1$ bonds) and $z_2$ of 
next-nearest-neighbors ($J_2$ bonds). Consequently we have 
$\alpha_c=1/2$
for the square
lattice but $\alpha_c=2/3$ for the bcc lattice. 
Therefore we define as the appropriate parameter of frustration 
\begin{eqnarray}
p =\frac{J_2 z_{2}}{J_1 z_{1}}
\end{eqnarray}
and we have $p_c^{square}= p_c^{bcc}=1/2$. 

In what follows we use the exact
diagonalization scheme and the linear spin wave theory
 to calculate the ground-state properties 
of the $J_1-J_2$ model on the bcc lattice and compare the
results with the corresponding ones for the square lattice.
We will present the ground-state energy, the violation of the Marshall-Peierls
sign rule, the sublattice magnetizations and the spin gap of finite lattices
in section II. Properties of the infinite lattices will be given in section
III, where the results of the extrapolation of the ground-state energy, the
sublattice magnetizations (from ED data) and the corresponding results of the
LSWT are shown.  

\section{Exact Diagonalization \label{ed}}
\subsection{The generation of finite bcc lattices}
The generation of finite 3d lattices with periodic boundary conditions 
is less transparent than for 2d lattices. As has been recently pointed 
out by Betts
and coworkers \cite{stewart97,Betts_Schulenburg98, Schulenburg_Flynn01}  
the use of a triple of edge vectors in upper triangular lattice form (utlf)
\cite{lyness} leads to a systematic generation
of  finite 3d lattices.
In this paper we use the utlf edge vectors and 
follow strictly  Refs.     
\cite{Betts_Schulenburg98, Schulenburg_Flynn01}. 
Finite parallelepipeds that build up the infinite bcc lattice can be
  defined by three edge vectors,
\begin{eqnarray}
{\bf L}_{\alpha}=\sum^3_{\beta =1}n_{\alpha \beta}{\bf a}_{\beta},
\end{eqnarray}

\noindent
where $n_{\alpha \beta}$ with $\beta =1,2,3$ are integers and ${\bf
  a}_1 =(1,1,-1)$, ${\bf a}_2 =(1,-1,1)$, ${\bf a}_3 =(-1,1,1)$ 
 are the basis vectors of the lattice connecting nearest neighbors. 
The lattice vectors
connecting next nearest neighbors are ${\bf b}_1 =(\pm 2,0,0), {\bf b}_2
=(\pm 0,2,0),
  {\bf b}_3 =(0,0,\pm 2)$.

There are altogether 
10 finite bcc
lattices with $N\leq 36$ listed in Table \ref{table1}, which fulfill the
following three conditions: 
(i) Every site $i$ of the bcc lattices should have 8 nearest and 6
next-nearest neighbors, which means that
they have the full number of nearest and next-nearest neighbors. 
(ii) The  finite lattices should 
be bi-bipartit in order to avoid frustration due to boundary conditions
for $p =0$ and $p
\longrightarrow \infty$. (Notice, that finite bcc lattices may be not
bi-bipartite even though the infinite bcc lattice is.)
(iii) Furthermore they should be topologically
distinct, i.e. the spin Hamiltonian (\ref{ham}) should  
exhibit different physical properties. 
\begin{table}
\caption{The ten finite bcc lattices are used for exact diagonalization.
${\bf L}_1$, ${\bf L}_2$, ${\bf L}_3$ are the three 
edge vectors in upper triangle lattice form.
$N$ is the Number of sites and $\gamma
=A,B,C,D,...$ is an additional label corresponding 
to a notation used in  \cite{Betts_Schulenburg98} to  
distinguish finite lattices with identical $N$. 
\label{table1}}
\begin{ruledtabular}
\begin{tabular*}
{\hsize}{c@{\extracolsep{0ptplus1fil}}c@{\extracolsep{0ptplus1fil}}cc}
 &\multicolumn{3}{c}{edge vectors}\\
N$\gamma$&${\bf L}_1$&${\bf L}_2$&${\bf L}_3$\\
\colrule
24C &(2,0,10)&(0,2, 6)&(0,0,24)\\
28D &(2,0,10)&(0,2, 6)&(0,0,28)\\
32D &(2,2, 4)&(0,8, 0)&(0,0, 8)\\
32F &(2,0, 6)&(0,4, 8)&(0,0,16)\\
32H &(2,0,10)&(0,2, 6)&(0,0,32)\\
32J &(4,0, 4)&(0,4, 4)&(0,0, 8)\\
32K &(2,0, 6)&(0,4, 4)&(0,0,16)\\
36A &(2,0,10)&(0,2, 6)&(0,0,36)\\
36B &(2,0,14)&(0,2,10)&(0,0,36)\\
36C &(2,2, 4)&(0,6, 6)&(0,0,12)
\end{tabular*}
\end{ruledtabular}
\end{table}

\subsection{Ground-state energy}
The ground-state energy gives first insight 
in the nature of possible zero-temperature phase transitions. 
In the thermodynamic limit a kink in $E_0(p)$
(respectively a jump in the first derivative 
$dE_0/d p$) signals a first order transition, whereas a smooth 
$dE_0/d p$ is compatible with second order transitions. Furthermore
the maximum in $E_0(p)$
indicates the point of maximal frustration. 
The ground-state energy  for the classical
model consists of two straight lines $\;E_0^{clas}(p) =(p -1)N\;$
for $p \le 0.5\;$ and $\;E_0^{clas}(p) =-p N\;$
for $p \ge 0.5$ with a kink (maximum) at $p=0.5$.

 The quantum ground state $|\Psi_0\rangle$ is a singlet eigen state 
of total spin for all $J_2$. 
In analogy to the square lattice \cite{squa_ref8} $|\Psi_0\rangle$
of finite bcc lattices with $N \mod 8=0$ 
has the same translational symmetry
for small and large $J_2$ (${\bf k}_{gs}=(0,0,0)$), 
whereas the translational 
symmetry of $|\Psi_0\rangle$ for lattices with $N \mod 4=0$ but 
$N \mod 8 \ne 0$ is  different 
for small and large $J_2$.
The change of symmetry from {\bf ${\bf k}_{gs}=(0,0,0)$} 
to {\bf ${\bf k}_{gs}=(\pi ,\pi ,\pi )$} appears slightly right from the
maximum of the ground-state energy.

\begin{table}
\caption{The ground-state energy per site of the 10 considered finite bcc
lattices
 for different values of $J_2$ ($J_1=1$).
\label{table2}}
\begin{ruledtabular}
\begin{tabular*}
{\hsize}{c@{\extracolsep{0ptplus1fil}}c@{\extracolsep{0ptplus1fil}}cc}
 &\multicolumn{3}{c}{$E_0 /N$}\\
N$\alpha$&$J_2=0$&$J_2=0.7$&$J_2=1.3333$\\
\colrule
24C &-1.21305&-0.73883&-1.34537\\
28D &-1.20223&-0.73003&-1.32544\\
32D &-1.19572&-0.72851&-1.31225\\  
32F &-1.19512&-0.72744&-1.31217\\  
32H &-1.19474&-0.72708&-1.31189\\
32J &-1.19440&-0.74003&-1.31688\\        
32K &-1.19408&-0.73203&-1.31423\\
36A &-1.18953&-0.72119&-1.30021\\
36B &-1.19264&-0.72180&-1.30145\\
36C &-1.19278&-0.72248&-1.30149\\
\end{tabular*}
\end{ruledtabular}
\end{table}

\begin{figure}[ht]
\epsfig{file=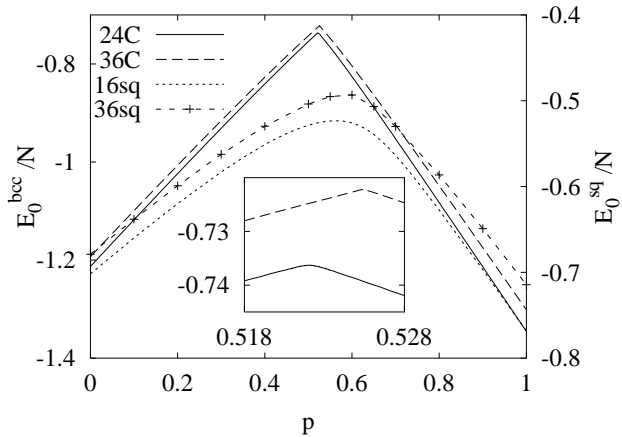, scale=0.67, angle=-0}
\caption{Ground-state energy per site in units of $J_1$ for the bcc lattices
  24C, 36C 
  and for the square lattices (scaled) 
  with $N=16$ (16sq) and $N=36$ (36sq) versus $p$. 
  The inset is an
  enlargement of the strongly frustrated region around $p =0.52$.}
\label{energies}
\end{figure}
   
For the quantum model we show exact-diagonalization results of $E_0/N$ for   
three different values of frustration in
Table \ref{table2}.
While $J_2=0$ and $J_2=1.3333$  correspond to zero or small frustration, 
$J_2=0.7$ is in the region of strong  frustration.

For the sake of clearness we present  in the figures only results
of selected finite lattices. To illustrate the finite-size effects 
in the most of the subsequent figures we 
present data  for the smallest (N=24) and the largest lattices we have
calculated,
where for $N=32,36$ we have chosen the lattices having   
highest symmetry.
Note, that the curves for 
lattices of identical $N$ look very similar.

For comparison with the square lattice we have 
recalculated data of \cite{squa_ref8} up to $N=6 \times 6 =
36$. 
However, we think a square lattice with $N=4 \times 4 =16$ 
is comparable to a bcc lattice with $N=36$.
This can be seen if one looks at the
characteristic lengths $L_{3d} \propto N^{1/3}$ and $L_{2d} \propto 
N^{1/2}$. Further, we mention that the square lattice of $N=16$
contains 5 neighborhood shells, whereas some of the more dense finite bcc
lattices with $N=32$ and $N=36$ contain even more.          

In Fig.\,\ref{energies} one  finds the
ground-state energies  of the bcc lattices 24C, 36C 
and for
comparison the $N=16$ and the $N=36$ square lattices.
To have comparable curves we scaled the ground-state
energy of the square lattice with the factor $7/4$.
Fig.\,\ref{energies} illustrates that, contrary to the 2d model, 
the ground-state energy of the 3d
bcc lattice
behaves very similarly to the classical model. As can be seen in 
Fig.\,\ref{energies} 
the kink in the ground-state energy is almost independent of 
the size of the bcc lattices. This can be interpreted as an indication that 
the kink survives in the thermodynamic limit.

\subsection{Ground-State Phase Relationships}

The phase relationships of the Ising basis states $|n\rangle$ in the ground
state $|\Psi_0\rangle $ of the bipartite Heisenberg antiferromagnet 
(i.e. $J_2=0$ in (\ref{ham}))  
follow the Marshall-Peierls sign rule \cite{Marshall}. This sign rule can be
formulated as 
\begin{equation}
\label{eigenstate}
|\Psi_0\rangle = \sum_n c_n |n\rangle ,\quad c_n >0.
\end{equation}
Here the Ising states $|n\rangle$ are defined by
\begin{equation}
|n\rangle \equiv (-1)^{N/2 -M(X)}|m_1 \rangle \otimes |m_2 \rangle
\otimes ... \otimes | m_N \rangle ,
\end{equation}
where $|m_i \rangle$, $i=1, ..., N$, are the eigenstates of the site spin
operator $S^z_i$ (i.e. $ m_i =\pm \frac{1}{2}$) and  
$M(X)=\sum_{i\epsilon X}m_i$. The standard Marshall-Peierls 
sign rule appropriate for
the N\'eel phase at small $p$ is obtained for $X=A$, i.e. 
$X$ labels one of the two equivalent  
sublattices.  For large values of $p$ we have antiferromagnetic order 
within the initial sublattices $A$ and $B$ and 
$A$ and $B$ resolve into
4 sublattices ($A \longrightarrow A_1$, $A_2$ and $B  \longrightarrow
B_1$, $B_2$). Then a modified sign rule holds
with $X=A_1 \cup B_1$.
 
As pointed out in \cite{zfp93,squa_ref9a} and very recently in 
\cite{squa_ref13a}
the sign rule may survive some frustration but is clearly violated 
for the square lattice in the strongly frustrated spin-liquid region. 
Hence we can use the violation of the Marshall-Peierls sign rule as 
an indication of the breakdown of the two-sublattice N\'eel state.

\begin{figure}[ht]
\epsfig{file=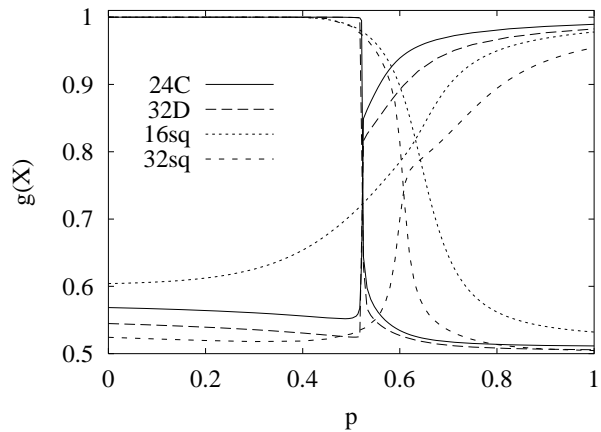, scale=0.67, angle=-0}
\caption{The weight $g(A)$ (that is 1 for 
$p=0$) and $g(A_1 \cup B_1)$ (that is 1 for $p \gg 1$)
 of basis states 
  $|n \rangle$
  fulfilling the Marshall-Peierls sign rule for the
  bcc lattice 24C,  
  36C and the
  square lattice with $N=16$ (16sq) and $N=32$ (32sq) 
  versus $p$. On the bcc lattices the
  sign rule is 
  completely fulfilled up to $p=0.4$ and 99.8\% fulfilled at $p=0.51$.}
\label{mrule}
\end{figure}

In Fig.\,\ref{mrule} one finds the weight {\bf
$g(X)=\sum'_n (c_n )^2$}   
of the Ising states $|n \rangle$ fulfilling the
Marshall-Peierls sign rule (i.e. the sum $\sum'_n{ }$ is restricted to the 
subset of states having $c_n > 0$) for two finite bcc lattices  and 
two square lattices. 
For the $J_1-J_2$ model on the bcc lattice the rule (\ref{eigenstate})
is violated 
almost discontinuously at that point where the ground-state 
energy has its maximum. 
This is a hint to a very drastic change of the ground state on the
bcc lattice around $p =0.52$ which can be attributed to an abrupt  
breakdown of the two-sublattice N\'eel state. On the other hand 
for the $J_1-J_2$ model on the square lattice the violation of the sign rule 
(\ref{eigenstate}) starts smoothly and becomes significant near $p=0.4$ where
the 
second-order transition to the spin-liquid state takes place.
The modified sign rule with $g(A_1 \cup B_1)$ changes also
discontinuously for the bcc lattice but smoothly for the square lattice.
However, there is a violation of the modified rule also in the
collinear phase, which can be attributed to the  
coupling between the both antiferromagnetic subsystems living on the 
initial sublattices $A$ and $B$.
Only for large $p \gg 1$, where the ground state becomes a product state 
of both antiferromagnetic subsystems  the modified  rule is 
rigorously fulfilled.

\begin{table}
\caption{Sublattice magnetizations $m^2$ and $m^2_{\alpha}$
of the 10 considered bcc lattices
 for different values of $J_2$. \label{tms}} 
\begin{ruledtabular}
\begin{tabular}{c@{\extracolsep{0ptplus1fil}}ccccc}
  &\multicolumn{1}{c}{$m^2$}
  &\multicolumn{1}{c}{$m^2_{\alpha}$}
\\
   N$\alpha$
  &\multicolumn{1}{c}{$J_2 =0$}
  &\multicolumn{1}{c}{$J_2 =1.3333$}
\\
\colrule
24C &0.09362&0.11006\\
28D &0.09057&0.10429\\
32D &0.08787&0.09958\\  
32F &0.08800&0.09959\\  
32H &0.08811&0.09964\\
32J &0.08819&0.09897\\        
32K &0.08827&0.09933\\
36A &0.08600&0.09605\\
36B &0.08516&0.09562\\
36C &0.08509&0.09561\\
\end{tabular}
\end{ruledtabular}
\label{table6}
\end{table}

\begin{figure}[bp]
\epsfig{file=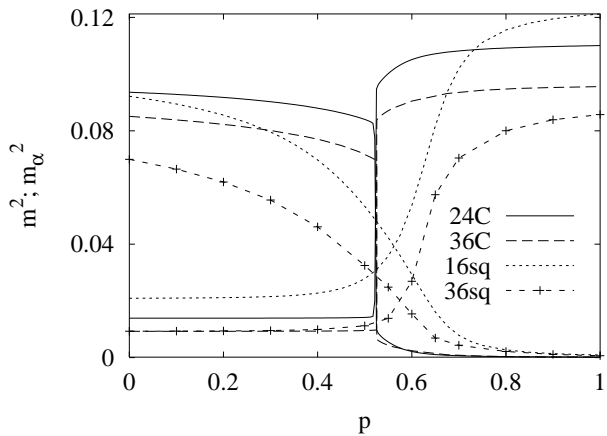, scale=0.67, angle=-0}
\caption{Sublattice magnetizations $m^2$ (maximal for small $p$) 
 and $m^2_{\alpha}$ (maximal for large $p$)
  of the bcc lattices 24C, 36C and the square
  lattices with $N=16$ (16sq) and $N=36$ (36sq) versus $p$. 
}
\label{fvms}
\end{figure}

\begin{figure}[bp]
\epsfig{file=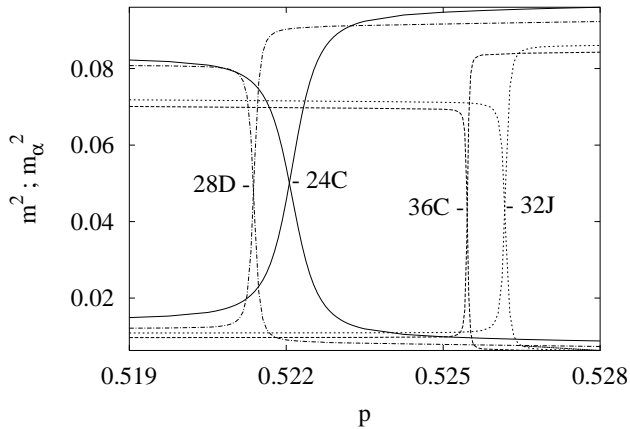, scale=0.67, angle=-0}
\caption{Sublattice magnetizations $m^2$ (maximal for small $p$)
 and $m^2_{\alpha}$ (maximal for large $p$)
  of the bcc lattice 36C, 32D, 28D and 24C near the phase transition point
  $p \approx 0.52$. 
}
\label{fms0.7}
\end{figure}

\subsection{Sublattice magnetizations}
Of course the most important parameter to study N\'eel ordering is the
sublattice magnetization. 
In finite systems the conventional antiferromagnetic LRO, that is realized for
small $p$, has to be described by the square of the sublattice 
magnetization
of one spin component
\begin{eqnarray}
\label{m1}
m^2 ({\bf Q})=
\langle [ \frac{1}{N}\sum^N_{i=1}e^{i{\bf Q}{\bf R}_i} S^z_i ]^2 \rangle 
\end{eqnarray}
with ${\bf Q}_{N\acute{e}el}=(\pi ,\pi , \pi)$ for the bcc lattice 
(and ${\bf Q}_{N\acute{e}el}=(\pi , \pi)$  
for the square lattice).
For large values of $p$ 
the magnetic wave vectors 
${\bf Q}^{1,2}_{col}=(\pm \pi /2 ,\pm \pi /2 ,\pm \pi /2)$
have to be used for the bcc lattice (and ${\bf Q}^{1}_{col}=(\pi ,0)$ or 
${\bf Q}^{2}_{col}=(0, \pi)$ for the square
lattice) to
describe the collinear phase with antiferromagnetic order within the
initial sublattices $A$ and $B$. We denote in the following 
the order parameter of the N\'eel phase calculated with 
${\bf Q}_{N\acute{e}el}$ with  $m^2$ and that of the  
collinear phase
calculated with ${\bf Q}^{1}_{col}$ or ${\bf Q}^{2}_{col}$ with $m^2_{\alpha}$.
Notice, that 
$m^2_{\alpha}$ is identical for ${\bf Q}^{1}_{col}$ and ${\bf
Q}^{2}_{col}$.
In Table \ref{tms} we give the order parameters of
the ten finite  bcc lattices for $J_2=0$ and $J_2=1.3333$. 
The behavior of $m^2$ and $m^2_{\alpha}$ 
shown in Fig.\,\ref{fvms} again illustrates, that the
influence of the frustration on the ground-state properties is basically
different for the square
and bcc lattice and suggests a direct first-order transition between both 
N\'eel phases for the bcc lattice.
A more detailed presentation of the transition region is given in 
Fig.\,\ref{fms0.7}. One finds that the position of the transition only 
slightly depends on size and symmetry of the finite lattices. Moreover, 
the width of transition region is getting smaller
with growing $N$. Again we mention that the region of transition
is related to the maximum of $E_0$ 
and the significant violation of the Marshall-Peierls
sign rule.

\begin{figure}[bp]
\epsfig{file=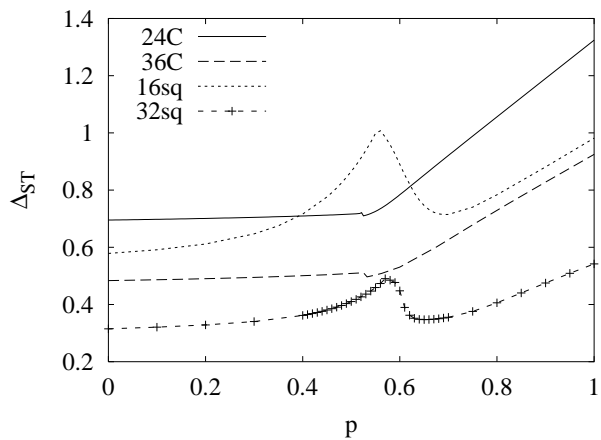, scale=0.67, angle=-0}
\caption{Spin gap $\Delta_{ST}$ in units of $J_1$, 
  i.e. the gap to the first triplet excitation
  for the bcc lattices 24C, 36C and the square lattices with N=16 (16sq) and 
  N=32 (32sq)
  versus $p$. \label{fsgap}}
\label{ms0.7}
\end{figure}

\subsection{Spin gap}
Another indication for a possible quantum disordered spin-liquid state is  
the spin gap, i.e. the gap $\Delta_{ST}$ 
between the singlet  ground state and  
the first triplet excitation. In N\'eel 
ordered systems we have Goldstone modes and no spin gap is observed in the
thermodynamic limit. Contrary to this, quantum disorder is accompanied by the
opening of a spin gap. We show the gap $\Delta_{ST}$ of two finite 
bcc lattices
  in Fig.\,\ref{fsgap}. 
  The first triplet excitation relevant for the gap belongs to the
translational symmetry
  ${\bf k}_{t}=(\pi , \pi , \pi)$ for small $p$ 
  and ${\bf k}_{t}=(\pi /2, \pi /2, \pi /2)$ for large $p$. 
  For comparison we  show the gap for the square
  lattice of $N=16$ and   $N=32$ sites.
Of course, the gap of a finite lattice is finite. However, in the 
long-range ordered N\'eel and collinear phases 
the extrapolation to the thermodynamic limit yields a vanishing gap.
Obviously there is no increase in the gap for the bcc lattices 
till about $p \approx 0.52$ where the
transition from the two-sublattice N\'eel phase to the collinear phase takes 
place. In the
collinear phase the relevant coupling parameter for excitations is $J_2$
instead of $J_1$ and consequently the gap increases linearly with $p$.
Clearly we see that there is no special behavior of $\Delta_{ST}$ near the
transition point $p \approx 0.52$.  
The gap for the square lattice shows a similar behavior for parameter regions
where magnetic long-range order in the ground state is present, but around
$p=0.5$ the behavior of $\Delta_{ST}$ is in contrast to the bcc
lattice. In this region  a quantum disordered gapped phase 
for the square lattice
is expected which is consistent with the significant increase of 
$\Delta_{ST}$ near $p=0.5$.   

We conclude from  the examination of the spin gap that there are no
indications for a quantum disordered gapped phase for the bcc lattice.

\section{Infinite bcc lattices}
\subsection{Finite-Size Extrapolation}
To obtain properties of the infinite bcc lattice we extrapolate the ED data
of all ten  lattices listed in Table \ref{table1}. The finite-size
extrapolation is a well elaborated approximation scheme successfully applied
to many 2d quantum spin systems like the $J_1-J_2$ model on the square
lattice \cite{squa_ref8}. But even for 3d lattices this scheme may lead to
precise data for the infinite lattice 
\cite{stewart97,Betts_Schulenburg98,Schulenburg_Flynn01}.
The corresponding scaling laws are known from literature
\cite{Hasenfratz_ZPB,Neuberger_PRB,Oitmaa_PRB}. 
The scaling equation for ground-state energy per site $\epsilon =E_0/N$ 
of the bcc lattice is 
\begin{eqnarray}
\epsilon(L)=\epsilon (\infty )+A_4 L^{-4}+ \ldots
\end{eqnarray} 
and for the 
order parameter
\begin{eqnarray}
m^2 (L)=m^2 (\infty )+B_2 L^{-2}+ \dots 
\end{eqnarray}   
with $L=N^{1/3}$.

The same relation is valid for $m^2_{\alpha}$. 
The results are presented in Figs.\,\ref{finfgs} and \ref{finfms}.  
The discussion of the data is given below.

\begin{figure}[bp]
\epsfig{file=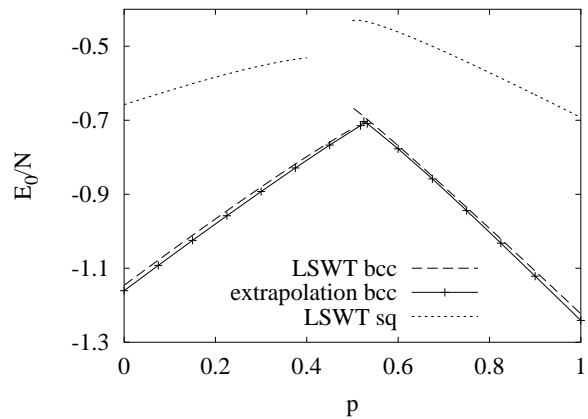, scale=0.65, angle=-0}
\caption{LSWT data for the ground-state energies $E_0$ of the 
infinite bcc lattice
  (dashed line, LSWT bcc) and  square lattice (dotted line, LSWT sq) 
as well as the
extrapolated ED data for $E_0$ (solid
  line with data points)
  versus $p$.\label{finfgs}} 
\end{figure}

\begin{figure}[bp]
\epsfig{file=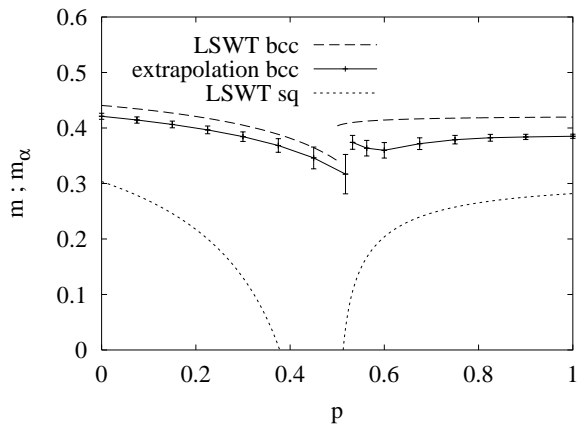, scale=0.65, angle=-0}
\caption{LSWT data for the sublattice magnetizations $m$ 
of the infinite bcc lattice (dashed
  line, LSWT bcc) and square lattice (dotted line, LSWT sq) 
as well as  extrapolated ED data
for  
   $\sqrt{3m^2 (\infty
  )}$  and  $\sqrt{3m^2_{\alpha} (\infty)}$ 
  (solid line with data points) vs. $p$. The error bars indicate the
standard deviation of the finite-size extrapolation.
\label{finfms}}
\end{figure}

\subsection{Linear spin wave theory (LSWT)}
\subsubsection{LSWT for small $J_2$}
Starting from the classical two-sublattice  N\'eel state 
we choose a two-boson representation of the Hamiltonian
(\ref{ham}). Rewriting the spin operators of the Hamiltonian in terms of
bose operators by using the usual Holstein-Primakoff transformation and
taking into account
only 
quadratic terms in the bose operators, we obtain 
a bosonic  Hamiltonian  in Fourier
transformed representation 
\begin{eqnarray}
\label{hfour}
H&=&s^2 N(-8+6p )+\sum_{\bf k} \{
A_{\bf k}(a^+_{\bf k}a_{\bf k}+ b^+_{\bf k}b_{\bf k})\\
&&\mbox{}+B_{\bf k}(a_{\bf k}b_{\bf -k}+ a^+_{\bf k}b^+_{\bf -k}) \}
\nonumber
\end{eqnarray}
with the coefficients
\begin{eqnarray}
A_{\bf k}&=&s(8-6p (1-\gamma_{2{\bf k}}))\\
B_{\bf k}&=&8s\gamma_{1{\bf k}}.
\end{eqnarray}
The structure factors of nearest neighbors and next-nearest 
neighbors are given by
\begin{eqnarray}
\gamma_{1{\bf k}}&=&\cos k_x \cos k_y \cos k_z \\
\gamma_{2{\bf k}}&=&\frac{1}{3}(\cos 2k_x +\cos 2k_y +\cos 2k_z).
\end{eqnarray}
The ground-state
energy per site is then
\begin{eqnarray}
\label{lswtgs}
E_0 /N=s^2(-8 +6p )+\frac{1}{N} \sum_{\bf k}(\omega_{\bf k}-A_{\bf k})
\end{eqnarray}
with $\omega_{\bf k}=\sqrt{A^2_{\bf k}-B^2_{\bf k}}$. The sublattice
magnetization $m=\langle S^z_{i}\rangle$ is
\begin{eqnarray}
m=s-\frac{1}{N}\sum_{\bf k}\left( -\frac{1}{2}
+\frac{A_{\bf k}}{2\omega_{\bf k}}\right) .
\end{eqnarray}
\subsubsection{LSWT for large $J_2$}
The classical ground state of the $J_1 -J_2$ model on the bcc lattice for
large $J_2$ consists of two interpenetrating N\'eel states  each living 
on the initial sublattices $A$ and $B$. The two N\'eel states are energetically
decoupled, i.e. the angle $\theta$ between 
the staggered magnetization on $A$ and $B$
is arbitrary for classical spins.  For the quantum model 
we start with arbitrary $\theta$ and use as quantization axis the
local orientation of the spins in the classical ground state. 
The further procedure is the same as for small $J_2$ but the bosonic
Hamiltonian now contains the angle $\theta$. By means of the
Hellmann-Feynman theorem   
\cite{Feynman_PR} 
$\langle
\partial H(\theta )/\partial \theta
\rangle 
=\partial E(\theta )/\partial \theta$ it can be easily found that in 
the quantum
model the collinear state ($\theta=0$ or $\pi$) has lowest energy.
This lifting of the continuous degeneracy of the classical
ground state by quantum fluctuations 
('{\it  order from disorder effect }') is also found for the 
square lattice \cite{Kubo_JPSJ}. 
For $\theta =0$ the bosonic Hamiltonian reads 
\begin{eqnarray}
H&=&-6Np s^2 +\sum_{\bf k} \bigg\{ A(a^{\dag}_{\bf k}a_{\bf
  k}+b^{\dag}_{\bf 
  k}b_{\bf k}) \\
 &+& 
[ C_{\bf k}(b_{\bf k}a^{+}_{\bf k}-b^+_{\bf k}a^+_{\bf
  -k}) -B_{\bf k}(a_{\bf k}a_{\bf -k}+b_{\bf k}b_{\bf -k})+h.c.]\bigg\}\nonumber 
\end{eqnarray}
where
\begin{eqnarray}
A&=&6sp\\
B_{\bf k}&=&3sp \gamma_{2{\bf k}}\\
C_{\bf k}&=&4s(\cos k_x \cos k_y \cos k_z \\
&&\mbox{}+i\sin k_x \sin k_y \sin k_z ).\nonumber
\end{eqnarray}
Then the ground-state energy per site is
\begin{eqnarray}
E_0 /N=-6s^2p +\frac{1}{N}\sum_{\bf k}\left( \frac{\omega_{1{\bf k}}}{2}+\frac{\omega_{2{\bf k}}}{2}-A\right)
\end{eqnarray}
with the modes 
\begin{eqnarray}
\omega_{1{\bf k}}= \sqrt{A^2 -4B^2_{\bf k}
+F_{\bf k}} \; ;\;
\omega_{2{\bf k}}= \sqrt{A^2 -4B^2_{\bf k}
-F_{\bf k}} 
\end{eqnarray}
and the function
\begin{eqnarray}
F_{\bf k}&=&
\left\{(C^2_{\bf k} -C^{*2}_{\bf k})^2 -8AB_{\bf k}(C^2_{\bf k}+C^{*2}_{\bf
    k})\right .\\
&&+ \left . 4C_{\bf k}C^*_{\bf k}(A^2 +4B^2_{\bf k})\right\}^{1/2}.
\nonumber
\end{eqnarray}
The sublattice magnetization is written as 
\begin{eqnarray}
m_{\alpha}=s&&- \, \frac{1}{N}\sum_{\bf k}
\frac{D({\bf k},\omega_{1{\bf k}})}
 {2\omega_{1{\bf k}}(\omega^2_{1{\bf k}}-\omega^2_{2{\bf k}})}\\
&&-\;\frac{1}{N}\sum_{\bf k}\frac{D({\bf k},\omega_{2{\bf k}}
  )}{2\omega_{2{\bf k}}(\omega^2_{2{\bf k}}-\omega^2_{1{\bf k}})} \nonumber
\end{eqnarray}
with 
\begin{eqnarray}
&&D({\bf k},\omega_{\bf k})=-\omega^3_{\bf k}+A\omega^2_{\bf k}
-(4B^2_{\bf k} -A^2)\omega_{\bf k}
\quad \quad \\
&& \quad +A(4B^2_{\bf k} +2C_{\bf k}C^*_{\bf k} )-2B_{\bf k}(C^2_{\bf k} 
+C^{*2}_{\bf k})-A^3 . \nonumber
\end{eqnarray}

The results of LSWT and the finite-size extrapolation are shown in 
Figs.\,\ref{finfgs} and \ref{finfms}. For the limits $J_2 =0$ and $J_1 =0$
our LSWT results are in agreement with data for the bcc and the sc lattice
given in \cite{Oitmaa_PRB}. Both methods yield 
similar results. For the ground-state energy we have a good quantitative
agreement. 
Being in the size of the data points, the standard deviation of extrapolated
ground-state energy is not shown.
For the order parameter the finite-size effects are stronger and
the agreement is only qualitative. Both methods suggest a first-order
transition for the spin-$\frac{1}{2}$ $J_1$-$J_2$ model on the
bcc lattice. The transition point obtained from the ED data is $J_2 \approx
0.7J_1$ (i.e. $p \approx 0.52$) while the LSWT becomes instable at the
classical transition point. 
     
\section{Conclusion}
We have presented spin-wave and exact diagonalization results for the 
spin-$\frac{1}{2}$ $J_1$-$J_2$ model on the
bcc  lattice and compare them with those for the square lattice. 
In general, we observe that the physics for the 3d quantum model is closer
to classical behavior since quantum fluctuations and finite-size corrections
become less important 
for higher coordination number and larger dimension.

We are not sure whether the increase of the magnetization $m_{\alpha}$ 
approaching the transition point from the right (shown in Fig.\,\ref{finfms}) 
is a real effect. A possible physical origin for an increase may be a stronger
coupling of the N\'eel ordered subsystems A and B 
due to larger quantum fluctuations or finite-size effects that
become more important in the region of strong frustration.
From the data for the  ground-state energy, the  Marshall-Peierls
sign rule, the sublattice magnetizations and the spin gap we conclude 
that the increase from dimension d=2 to d=3 
changes the physical properties basically. 
The good agreement with the spin wave results support this conclusion.
Contrary to the 2d model, where the quantum fluctuations and frustration 
lead to a second order transition from the  two-sublattice N\'eel state to  
a disordered spin-liquid like  
phase, in the 3d model
we find no indications for a disordered ground-state phase.
The quantum $J_1-J_2$ model on the bcc lattice shows one transition of first 
order induced by strong frustration from the  two-sublattice N\'eel state 
directly 
to the collinear state, where the transition takes place at $\alpha_c=
(J_2/J_1)_c \approx 0.7$. \\

{\bf Acknowledgements:}
We acknowledge support from the Deutsche Forschungsgemeinschaft
(Project No.  Ri 615/7-1). Besides 
we also want to thank Dirk Schmalfu{\ss} for fruitful
discussions. 



\begin{thebibliography}{99}
\bibitem{squa_ref7} P.~Chandra and B.~Doucot,  {\it Phys. Rev. B} {\bf 38},
  9335 (1988).
\bibitem{squa_ref8} H.J.~Schulz and T.A.L.~Ziman, {\it Europhys. Lett.}
  {\bf 18}, 355 (1992);
H.J. Schulz, T.A.L.~Ziman and D.~Poilblanc {\it J.~Phys.~I}
  {\bf 6}, 675 (1996).
\bibitem{squa_ref9} 
   J.~Richter, Phys. Rev. B $\bf47$, 5794 (1993).
\bibitem{zfp93}
K.~Retzlaff,  J.~Richter and  N.B.~Ivanov,
      Z. Phys. B {\bf 93}, 21, (1993).
\bibitem{squa_ref9a} 
 J.~Richter, N.B.~Ivanov and K.~Retzlaff,
      Europhys. Lett. {\bf 25}, 545 (1994).
\bibitem{squa_ref10} R.F.~Bishop, D.J.J.~Farnell and J.B.~Parkinson,
 {\it  Phys. Rev. B} {\bf 58}, 6394 (1998).
\bibitem{squa_ref11} R.R.P.~Singh, Zheng Weihong, C.J.~Hamer and J.~Oitmaa,
{\it Phys. Rev. B} {\bf 60}, 7278 (1999).
\bibitem{squa_ref12} V.N.~Kotov and O.P.~Sushkov,
{\it Phys. Rev. B} {\bf 61}, 11820 (2000).
\bibitem{squa_ref13} L.~Capriotti and S.~Sorella, 
{\it Phys. Rev. Lett.} {\bf 84}, 3173 (2000).
\bibitem{Sushkov01}
    O.P.~Sushkov, J.~Oitmaa and Zheng Weihong,
   {\it Phys. Rev. B}
   {\bf  63}, 104420 (2001).
\bibitem{squa_ref13a} L.~Capriotti, F.~Becca, A.~Parola and S.~Sorella, 
{\it Phys. Rev. Lett.} {\bf 87}, 97201 (2001).
\bibitem{1d_a} K.~Okamoto and K.~Nomura, 
{\it Phys. Lett. A} {\bf 169}, 433 (1992).
\bibitem{1d_b} S.~Eggert,
{\it Phys. Rev. B} {\bf 54}, R9612 (1996).
\bibitem{1d_c} Steven~R.~White and Ian~Affleck,
{\it Phys. Rev. B} {\bf 54}, 9862 (1996).
\bibitem{canals98} B.~Canals and C.~Lacroix,
  Phys. Rev. Lett. {\bf 80}, 2933 (1998).
\bibitem{koga01} A.~Koga and N.~Kawakami, 
  Phys.Rev.B {\bf 63}, 144432 (2001).
\bibitem{stewart97} G.E.~Stewart, D.D.~Betts  and J.S.~Flynn 
{\it J. Phys. Soc. Jap.} {\bf 66} 3231 (1997).
\bibitem{Betts_Schulenburg98}
 D.D.~Betts, J.~Schulenburg, G.E.~Stewart, J.~Richter and 
 J.S.~Flynn, 
   {\it J. Phys. A: Math. Gen.}
   {\bf  31}, 7685 (1998).
\bibitem{Schulenburg_Flynn01}
    J.~Schulenburg, J.S.~Flynn, D.D.~Betts and J.~Richter, 
   {\it Eur. Phys. J. B}
   {\bf  21}, 191 (2001).
\bibitem{lyness}
J.N.~Lyness, T.~Sorevik  and P.~Keast  {\it Mathematics of
        Computation} {\bf 56} 243 (1991).
\bibitem{Marshall}
    W.~Marshall, 
   {\it Proc. Roy. Soc. A}
   {\bf  232}, 48 (1955).
\bibitem{Neuberger_PRB}
    H.~Neuberger and T.~Ziman, 
   {\it Phys. Rev. B}
   {\bf  39}, 2608  (1989).
\bibitem{Hasenfratz_ZPB}
    P.~Hasenfratz and F.~Niedermayer, 
   {\it Z. Phys. B}
   {\bf  92}, 91  (1993).
\bibitem{Oitmaa_PRB}
    J.~Oitmaa, C.J.~Hamer and Zheng Weihong, 
   {\it Phys. Rev. B}
   {\bf  50}, 3877  (1994).

\bibitem{Feynman_PR}
    R.P.~Feynman, 
   {\it Phys. Rev.}
   {\bf  50}, 340  (1939).
\bibitem{Kubo_JPSJ}
    K.~Kubo and T.~Kishi, 
   {\it J. Phys. Soc. Jap.}
   {\bf  60}, 567  (1990).
\end{thebibliography}
\end{document}